\documentclass[fleqn,twoside]{article}
\usepackage{espcrc2}

\usepackage{graphicx}
\usepackage{amssymb}

\title{Neutrino Astronomy in the Ice}

\author{T. Montaruli\address[UW]{Physics Department, 
        University of Wisconsin-Madison, 
        WI-53706, USA}%
        \thanks{on leave of absence from University of Bari, Physics Department and INFN, I-70126.}
   }
       
\begin{document}

\begin{abstract}
The South Pole is an optimal location for hosting astrophysical observatories. The status of the construction of the IceCube Observatory and some selected physics results will be discussed. Moreover prospects for detection of Ultra-High Energy cosmogenic neutrinos and techniques that can address this energy region will be considered.
\vspace{1pc}
\end{abstract}

\maketitle

\section{ICECUBE OBSERVATORY STATUS AND SELECTED RECENT RESULTS}
\label{sec1}

\begin{figure}[htb]
\includegraphics[width=17pc,height=18pc]{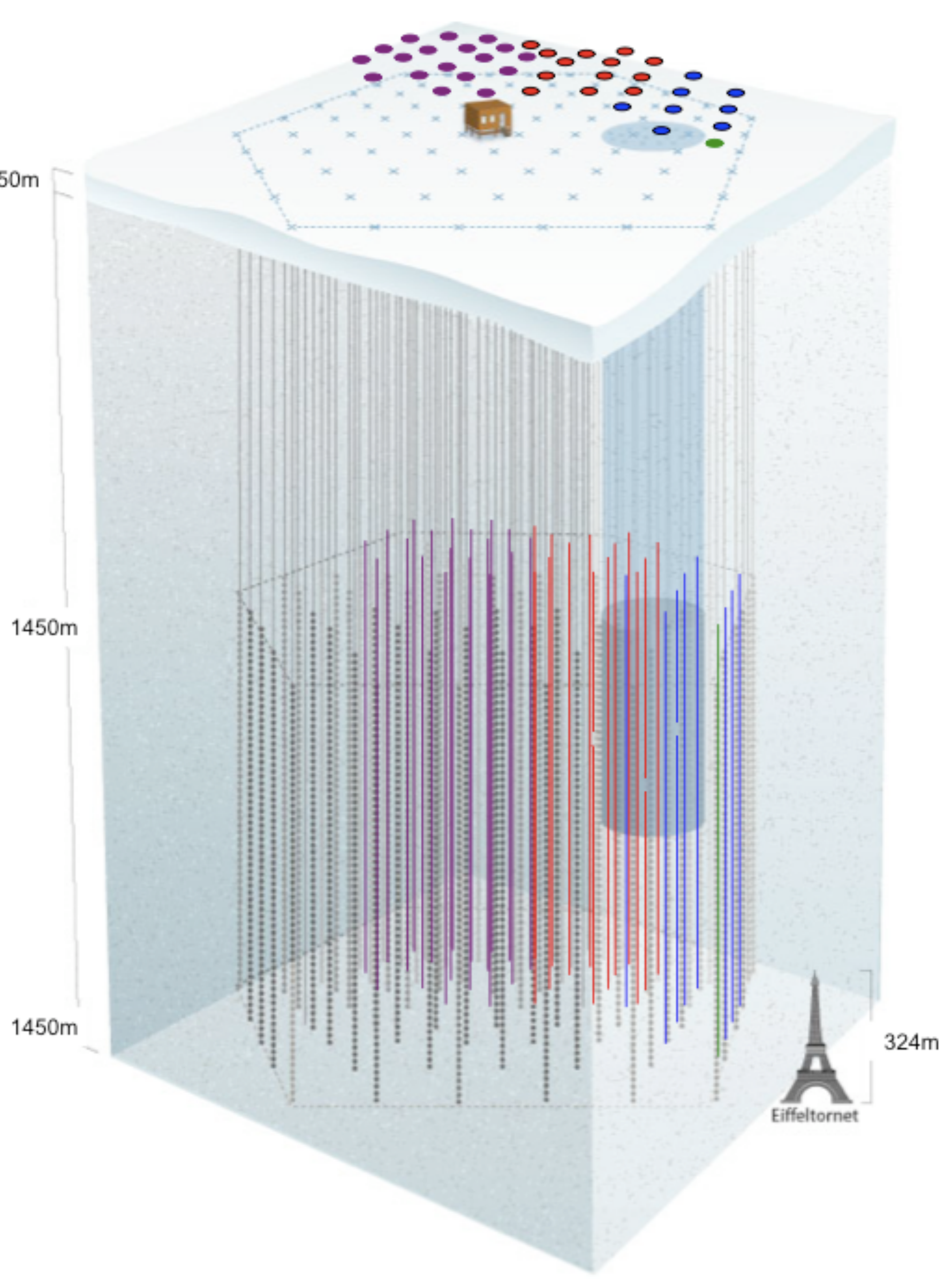}
\caption{Layout of the IceCube deep ice detector and surface array IceTop. Indicated (in different colors) are the stations and strings deployed in the 4 construction seasons from 2005 to the current one (Nov. 08 - Feb. 09). The small cylinder indicates the dimension of AMANDA with 19 strings holding 677 optical modules.}
\label{fig1}
\end{figure}

\begin{figure}[htb]
\includegraphics[width=18pc,height=18pc]{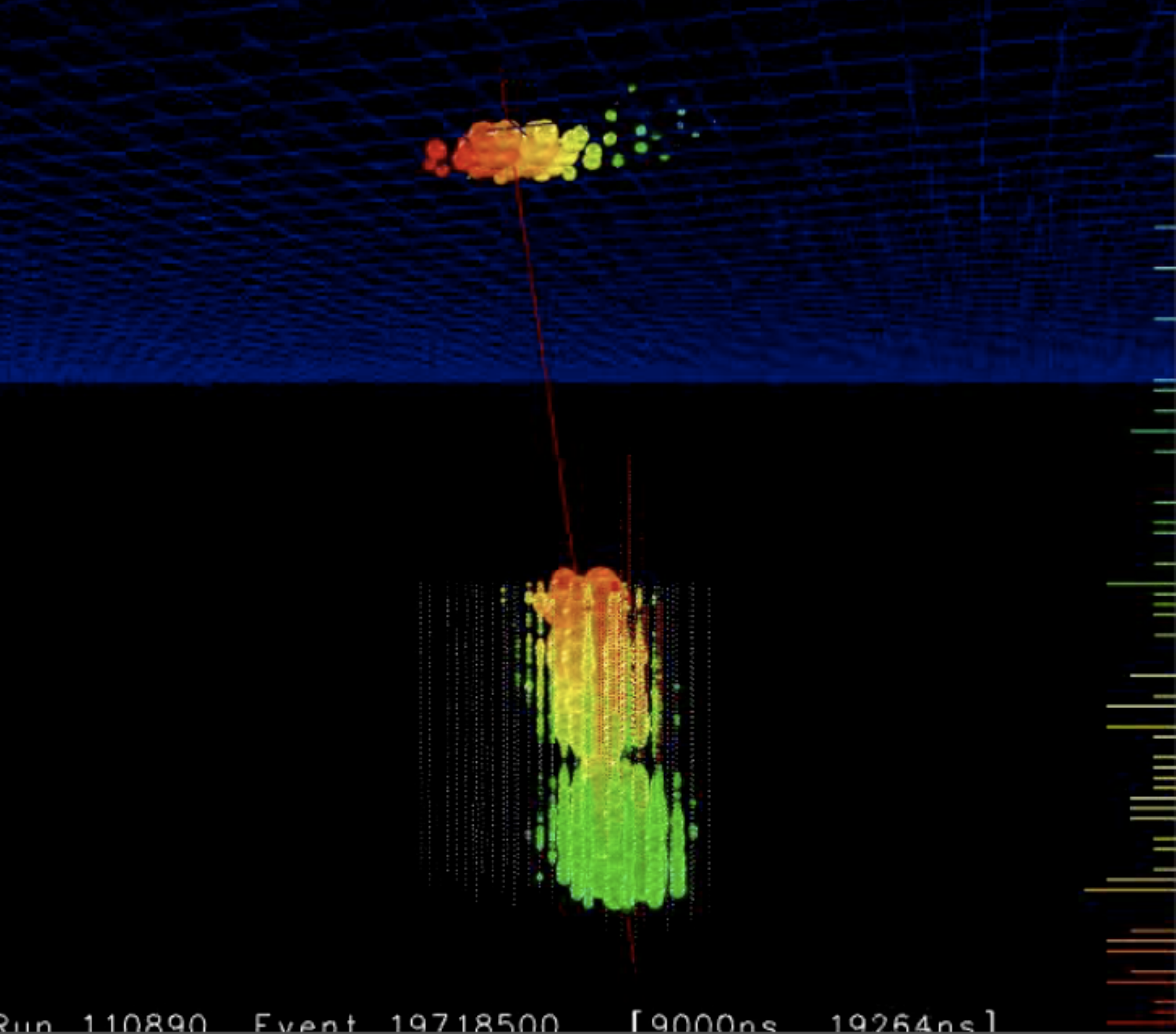}
\caption{A coincident event between IC40 and IceTop detected on Apr. 24, 2008, that released a total charge in IC40 of  $10^5$ photoelectrons.}
\label{fig2}
\end{figure}

The IceCube Observatory at the South Pole is composed of a Cherenkov radiation detector made of instrumented strings between about 1.5 and 2.5 km and a surface array called IceTop. Each string holds 60 digital optical modules (DOMs) separated vertically by about 17 m. Strings are about 125 m apart, as shown in Fig.~\ref{fig1}, where the configuration taking now data with 40 strings (IC40) and 40 IceTop stations is indicated. IceTop is made of stations corresponding to each string of two frozen water tanks. Each tank contains 2 DOMs that can measure the light emitted by the electromagnetic component of atmospheric showers and the combination of the 2 detectors can measure also the muons that penetrate deep in the ice. An example of such events with very high energy and a big bundle of muons penetrating deep in the ice is shown in Fig.~\ref{fig2}. IceCube construction will be concluded when 80 strings will be installed at a rate of about 16 per season up to the 2010-11 season. Construction within this schedule has been made possible thanks to an improved drilling system compared to the precursor detector AMANDA, at shallower depth between about 1.5 ad 2 km to be decommissioned next year. Holes 2.5 km long of about 60 cm in diameter are obtained using a hot water drill system; it takes about 40 hrs to drill them and about 12 hrs to lower the string of DOMs into one hole.

Additional 6 strings, called DeepCore, are planned to be installed with 50 out of 60 of high quantum efficiency sensors deployed with vertical spacing of 7m in the clear ice below 2100 m. The region of dense string spacing will consist of a total of 13 strings covering an area of roughly 250 m in diameter. The aim of this densely instrumented array is to lower the threshold of IceCube for enhancing sensitivity in the region below 1 TeV, interesting for dark matter studies, neutrino oscillations or galactic sources with steep spectra or cut-off at about a few TeV. The large instrumented volume of IceCube around DeepCore will offer a muon veto that will allow to single out starting muons in the detector due to low energy neutrino interactions in the fiducial volume.

Another extension under consideration is to enhance the effective area of IceCube for energies $>100$ TeV. A modified arrangement of the outer 12 strings can lead to an increase in effective area between 15-40\% for well reconstructed tracks depending on their distance from the rest of the array. This could be advantageous for the detection of cosmogenic neutrinos (see Sec.~\ref{sec:GZK}).

Currently 2400 deep and 160 surface DOMs have been deployed and less than 3\% are dead or have acquisition problems, most common ones affecting the local coincidence connections. This low percentage allows to estimate a survival rate after 15 yrs of about 95\% DOMs. Ice is in fact a very quiet environment for electronics operation and DOMs can be operated at a noise rate of 300 Hz on average. Such low rate compared to sea water, where rates are between about 30-70 kHz, makes supernova collapse searches possible exclusively in ice detectors. For instance IceCube expects a sensitivity of about 5$\sigma$
for a supernova with the luminosity of SN1987A located in tha Large Magellanic Cloud. Nonetheless, ice is a medium with larger scattering than sea water and it is depth dependent. Accounting for this dependence is a challenge and requires work to tune MC and data agreement through calibration tools (flashers or lasers) and muon time residuals (difference between measured hit times and time photons are expected to take to reach a DOM from the muon track position). 

DOMs are made of a 35 cm diameter pressure vessel (Benthosphere) containing  Hamamatsu R7081-02 photomultipliers with photocathode diameter of 25 cm. They have low power consumption of 3.5 W each and collect data autonomously and send them to surface. The DAQ system \cite{daq} includes a custom chip for waveform digitization at 300 MHz for 400 ns (ATWD) and a commercial 40 MHz Flash ADC that records 6.4 $\mu$s of data after each trigger. Data acquisition is launched by a discriminator with a threshold set at about 0.35 photoelectrons. Each DOM includes a precision clock synchronized every few seconds to a master clock at surface synchronized to GPS. DOMs have a self-calibration circuitry that sends a signal up and down through the cable and calibrations ensure a time resolution of about 2 ns \cite{first}.
\begin{figure}[htb]
\includegraphics[width=18pc,height=12pc]{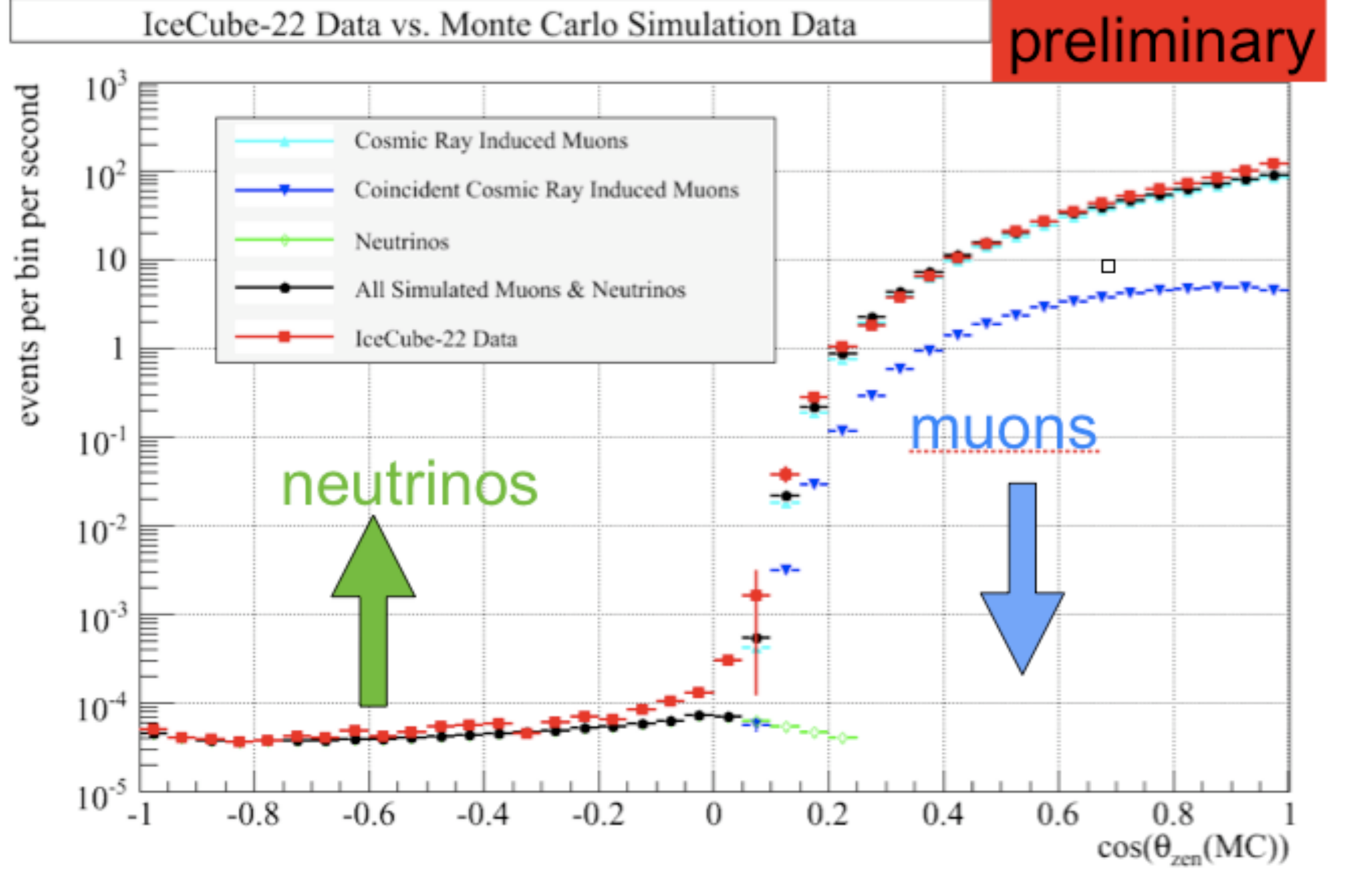}
\caption{Preliminary zenith angular distribution of muons from the nadir to the zenith for IC22 data (Muon filter merged to MB data - red squares). Errors are only statistical and rates are normalized to trigger level. Three contributions for the MC are summed (black full circles): atmospheric muons produced by the interaction of a cosmic ray in the atmosphere (cyan triangles), well reconstructed coincident muons produced by 2 cosmic rays interacting in the atmosphere and producing two independent showers in the readout window of $\pm 10\mu$s (blue inverted triangles) and neutrino induced upgoing muons (green diamonds).}
\label{fig3}
\end{figure}

Several hardware triggers are implemented and when a trigger condition is satisfied the system saves all hits in a readout window of $\pm 10 \mu$s. The main trigger is a simple majority trigger (SMT8) requiring 8 DOMs are hit within 5 $\mu$s. A single string trigger requires that 5 out of 7 adjacent DOMs are fired within 1.5 $\mu$s. Data at the South Pole are filtered and sent to the North by satellite on a bandwidth of about 37 GB/d for IC22. The SMT8 trigger rate determined by cosmic muons reaching the deep detector in IC22 is 550 Hz and in IC40 about 1 kHz. The muon filter, used for muon neutrino studies, selects upgoing muons from the lower hemisphere with a minimum requirement on the number of hit channels of 10 and higher energy muons above the horizon up to zenith of $50^o$ and its rate in IC40 is about 20 Hz. This filter, as well as others, selects data useful for different analysis that are further processed to higher levels with more refined reconstructions in the North and then used for analyses. We also select Minimum Bias (MB) data (one trigger every 200 in IC22). The physics run for IC22 started on May 31, 2007 and lasted up to Apr. 5, 2008 after which the IC40 run started. About 90\% of this period makes up the 275.7 d of livetime that was used after a good run selection for point source and atmospheric neutrino analyses. The down-time includes periods in which the detector is in an unstable configuration during the construction season.
Good runs are selected removing short runs of less than 0.5 hrs and based on a requirement of compatibility with a sliding average muon rate that accounts for the  $\pm 10\%$ seasonal variations at the South Pole.\begin{figure}[t]
\includegraphics[width=18pc,height=12pc]{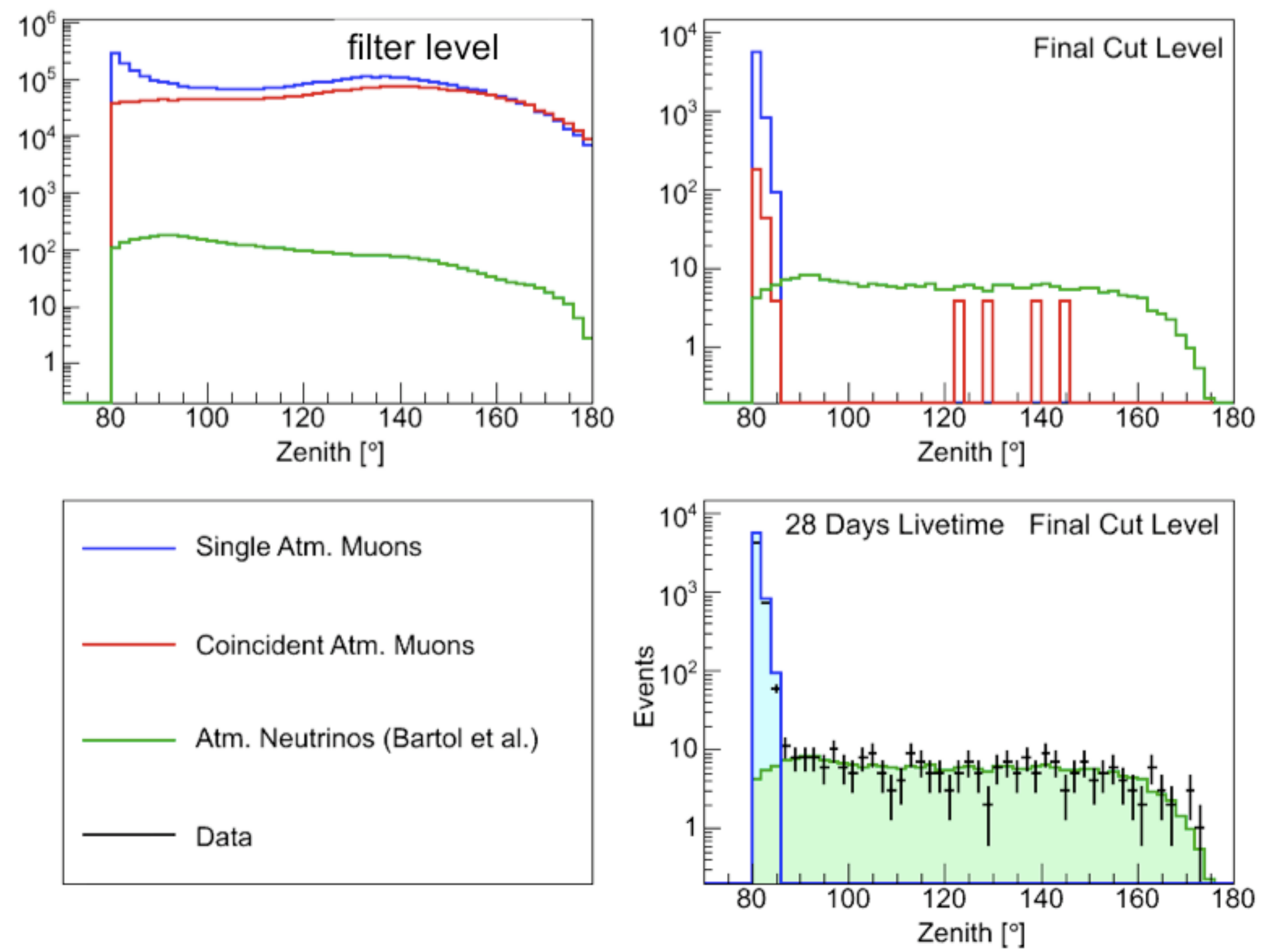}
\caption{In the top left panel the three MC contributions at filter level are shown. From top to bottom: muons produced by single CRs interacting in the atmosphere, coincident muons and atmospheric neutrinos. Clearly the region of upgoing muons is dominated by misreconstructed events at these level. In the top right panel the same three contributions (same order from top to bottom) are shown after all analysis cuts for the point source analysis that aim at achieving a good angular resolution are applied. In the bottom panel these are compared to 28 d of data showing a very good agreement.}
\label{fig4}
\end{figure}
\begin{figure}[t]
\includegraphics[width=18pc,height=12pc]{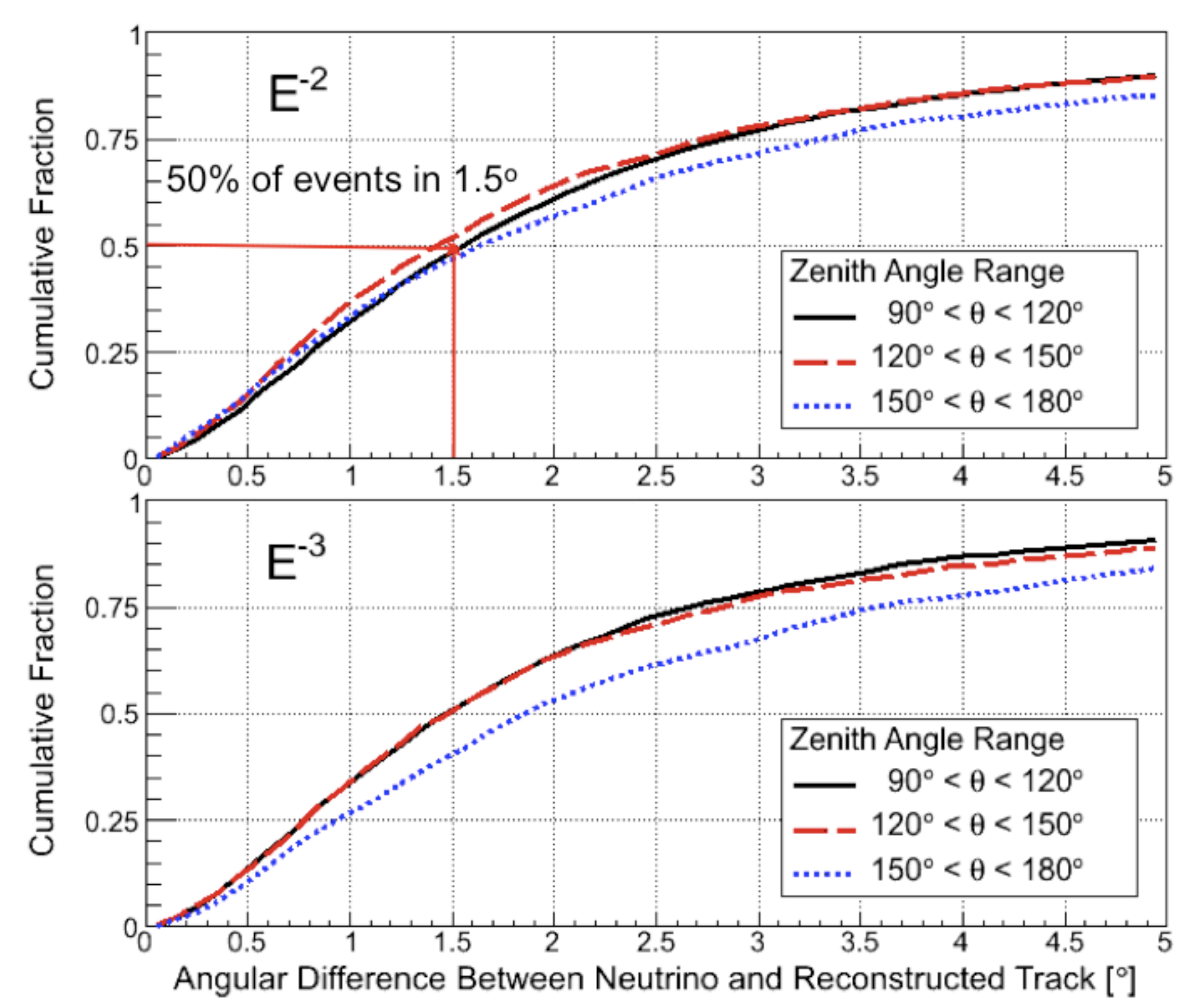}
\caption{Upper (lower) panel: PSF (fraction of events reconstructed inside an angular distance from the neutrino source direction) for an $E^{-2}$ ($E^{-3}$) flux of neutrinos producing 90\% of the upgoing muons in the range 3 TeV-3 PeV (250 GeV - 16 TeV). The PSF, shown in bins of zenith $\theta$, is mildly dependent on energy and declination $\delta$ (at South Pole $\delta = \theta - 90^{o}$).
For an $E^{-2}$ spectrum, 50\% of the events are included in a cone of angular radius of $1.5^{o}$.}
\label{fig5}
\end{figure}
\begin{figure}[htb]
\includegraphics[width=18pc,height=11pc]{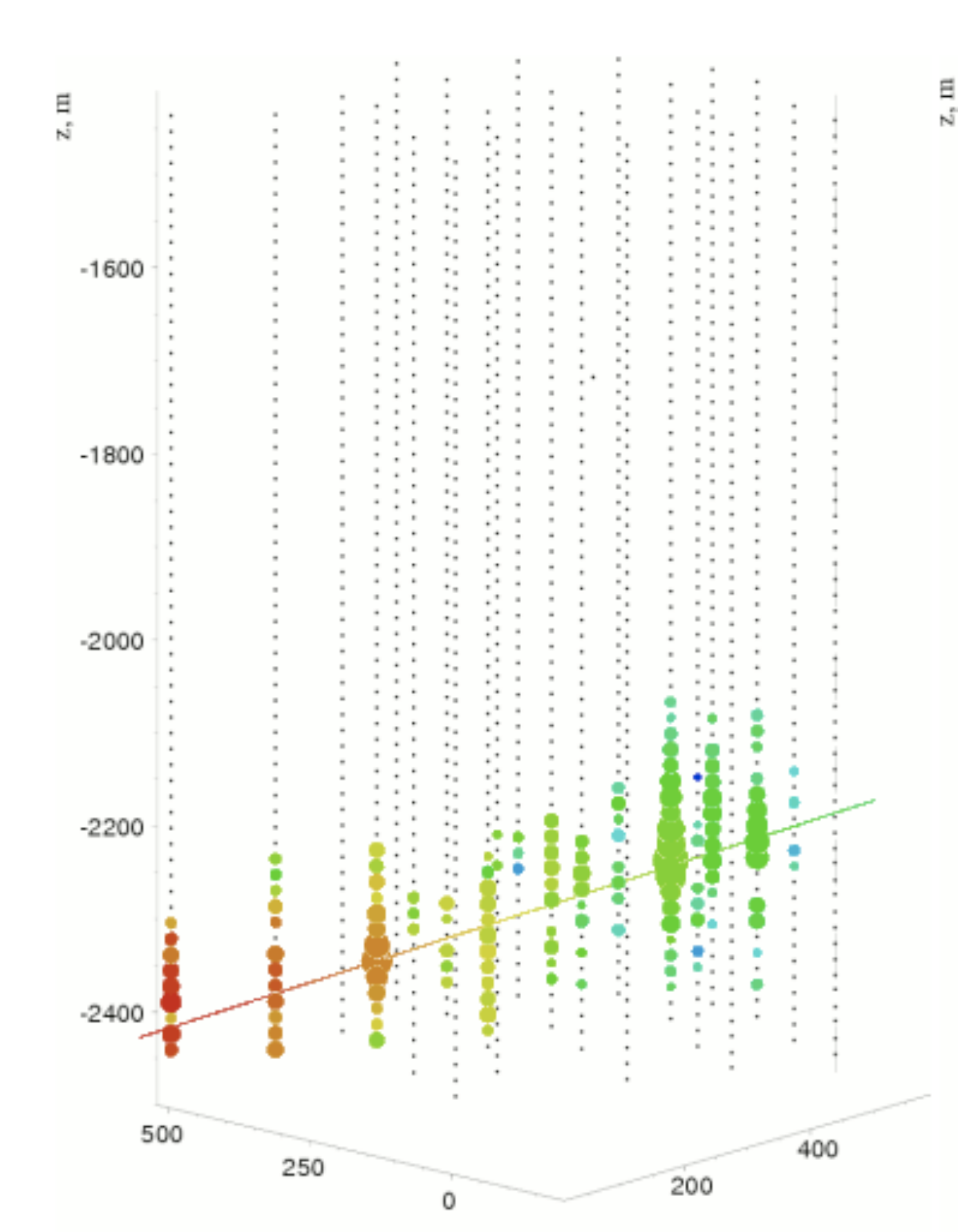}
\caption{One of the highest energy horizontal neutrinos in the hot spot in the IC22 point-source analysis. This event switched on NCh = 145 DOMs on 16 strings deep in the ice where the transparency of the ice allows photons to propagate hundreds of meters. From the MC of atmospheric neutrinos we expect about 2.3 neutrinos with NCh $> 140$ per year from the entire hemisphere and 0.4 from the horizontal declination band between $6^o-16^o$ where we measure 3. Neutrinos producing such high numbers of hit can have energies $\gtrsim 500$ TeV.  }
\label{fig6}
\end{figure}
\begin{figure}[t]
\includegraphics[width=20pc,height=15pc]{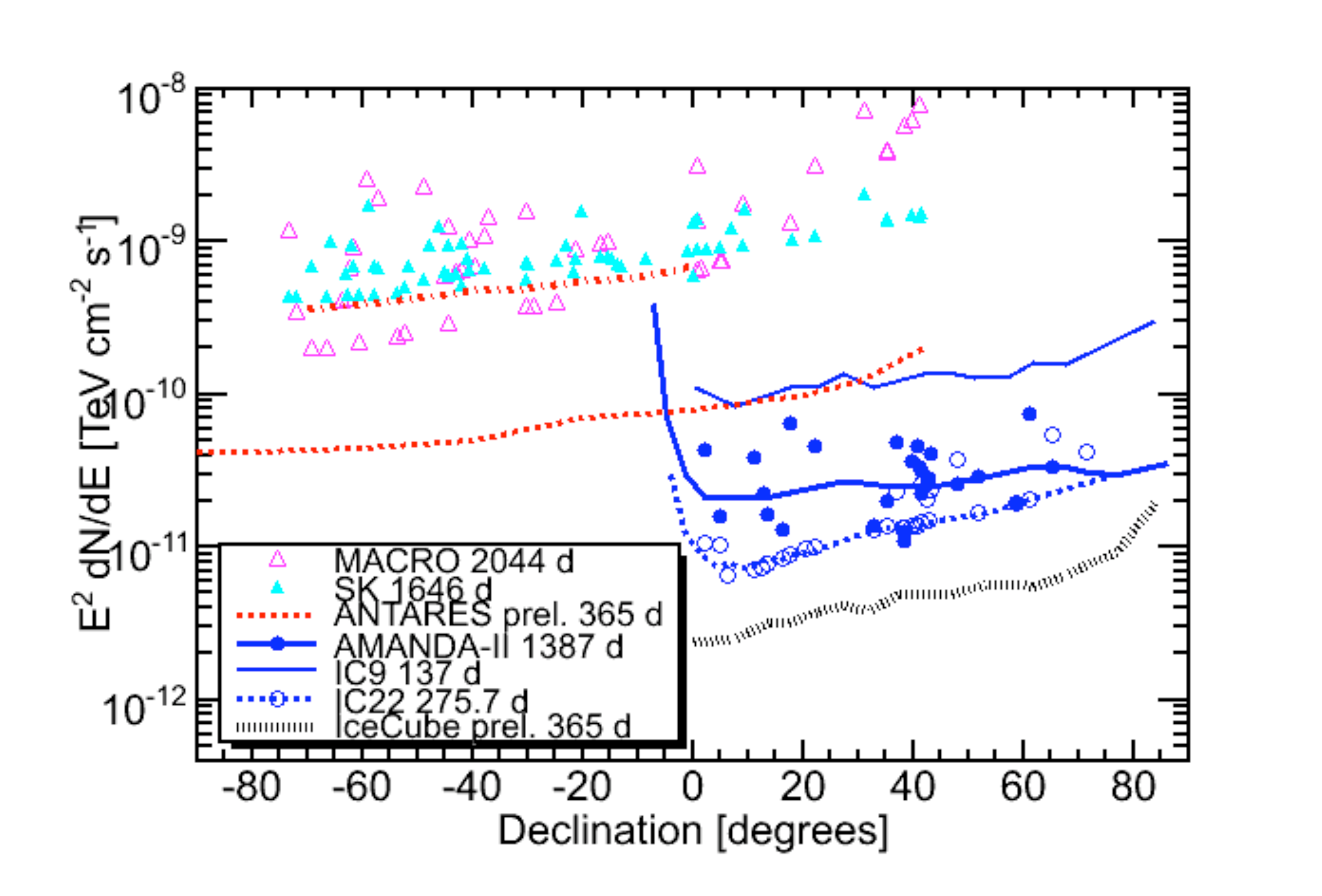}
\caption{Results for point-like source searches for $E^{-2} (\nu_{\mu} + \bar{\nu}_{\mu}$)  fluxes  vs declination. On the right from top to bottom (all 90\%cl): IC9 sensitivity for 137 d \protect\cite{taup07}, sensitivity and upper limits for specific sources (circles) for 7 yrs of AMANDA-II \protect\cite{amanda7} and 22 strings. Data for these configurations have been unblinded while the sensitivity for IC80 is from simulation and not optimized.
For the other hemisphere (on the left): sensitivity of 5 lines of ANTARES after 140 d. Predicted results  
for 1 yr of the full ANTARES (dashed line).  Upper limits for catalogues of selected  sources are shown for Super-Kamiokande 
(full triangles) \protect\cite{sk} and MACRO (empty triangles) \protect\cite{macro}.}
\label{fig7}
\end{figure}

The combination of the MB data and of the Muon filter has been analyzed to select well reconstructed muons with an uniform angular resolution in zenith of around $0.8^o$ (median). As an example, we show in Fig.~\ref{fig3} the full sky angular distribution of data compared to MC showing good agreement aside from the very horizontal region where the MC statistics is insufficient and the muon flux rapidly decreases due to the large ice layer muons have to cross. The aim of the plot is to show overall agreement between data and MC both for atmospheric muons and neutrinos. The reconstruction quality is good enough to see the transition from one component to the other, while the horizontal region (where astrophysical interest is for diffuse high energy neutrino fluxes) is being looked at with dedicated analyses. Cuts on the reconstruction quality, on the number of hit channels and on the amount of unscattered light hits are applied to extract a sample of neutrino induced upgoing muons suitable for point source analyses, described in these proceedings \cite{bazo}, containing a contamination of less than 10\% of misreconstructed atmospheric muons, as summarized in Fig.~\ref{fig4}. At this level of selections 5114 upgoing muon events have been selected and the agreement between data and MC is well inside the systematic uncertainty of atmospheric neutrino calculations of about 15\% \cite{bartol}. We expect 4642 atmospheric neutrinos using the Bartol flux calculation \cite{bartol} and a contamination of misreconstructed atmospheric muons of about 10\% of which a consistent part is due to coincident cosmic rays producing muons from two different directions. Also the achieved point spread function (PSF) is shown in Fig.~\ref{fig5}. A likelihood method was applied to look for excesses of high energy neutrino events clustered around any direction in the sky or around a catalogue of candidate sources on top of the atmospheric neutrino background. For this analysis we scramble in time measured events in declination bands in order to reproduce many background-only `equivalent experiments'. We find a hot spot in the all-sky search corresponding to a 1.3\% post trial probability (p-value) to be a fluctuation of the background. Though this p-value is not significant enough to claim any evidence of a neutrino source, enough data have already been collected with IC40 to verify or exclude this as a possible signal. This analysis also uses the energy information based on the fact that the astrophysical neutrinos are expected to have a much harder spectrum than atmospheric neutrino ones. In Fig.~\ref{fig6} we show one of the highest energy events that contributed most to the p-value of the hot spot region. A summary plot of the upper limits of AMANDA and IceCube and of other Northern hemisphere detectors is given in Fig.~\ref{fig7}.

\section{UHE NEUTRINO DETECTION}
\label{sec:GZK}

\begin{figure}[hbt]
\includegraphics[width=20pc,height=14pc]{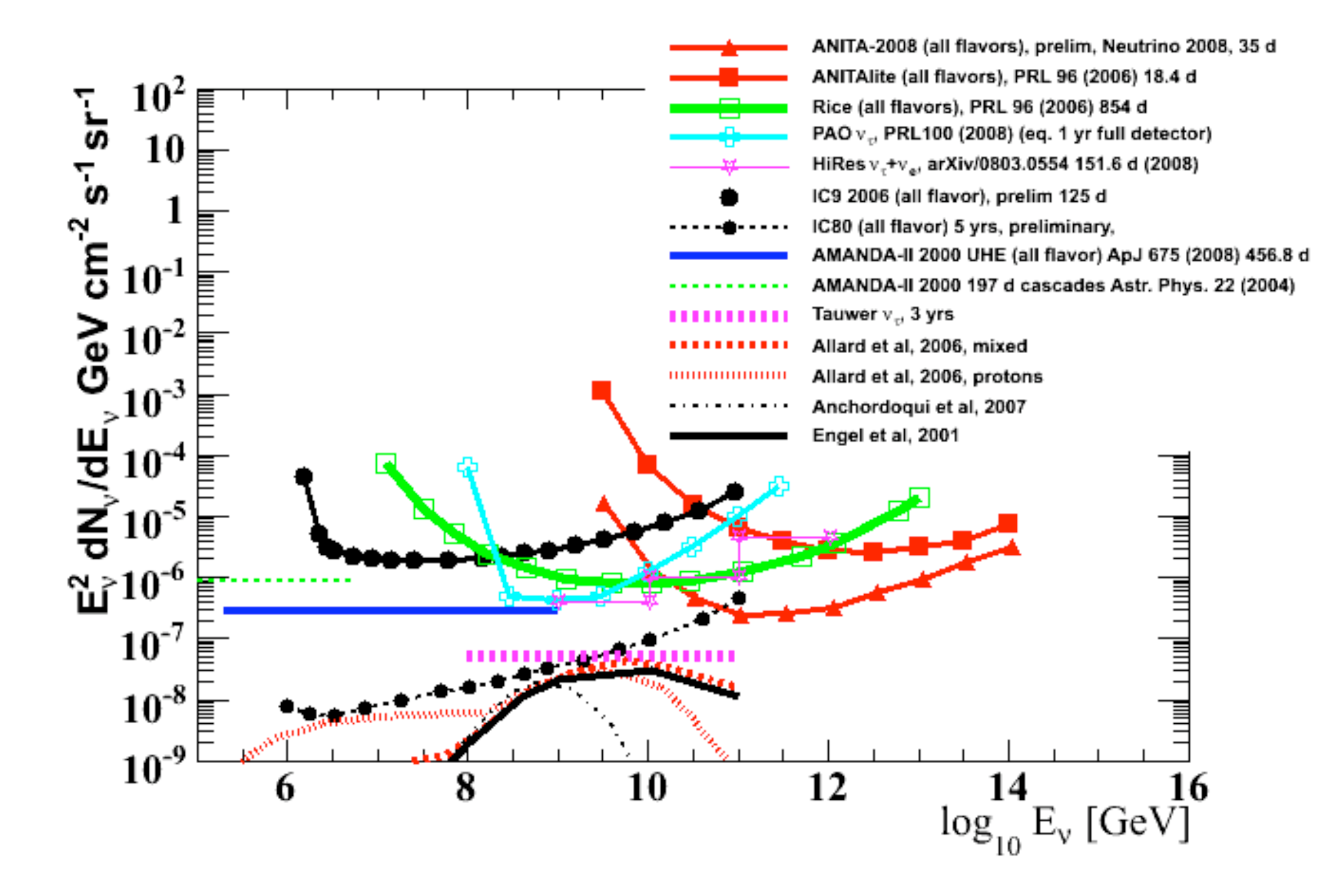}
\caption{90\% c.l. upper limits differential in energy on $\nu_{e}+\nu_{\mu}+\nu_{\tau}$ ($1:1:1$ oscillation assumption) for the ANITA radio detector prototype (red line-full squares) \protect\cite{anita}, the Dec 06-Jan 07 flight (red line-full triangles) \protect\cite{gorham} and Rice \protect\cite{rice} (scaled from 95\% c.l. to 90\% for 0 selected events - green open squares) sensitive to the EM component of $\nu$ induced showers in the ice sheet, 
IC9 upper limit (solid black line with full circles),  preliminary estimate for IC80 (dashed line with full circles) \protect\cite{eheicrc07}.
PAO limit is on the Earth-skimming $\nu_{\tau}$ flux using horizontal showers \protect\cite{PAO} and HiRes limit is on $\nu_e + \nu_{\tau}$ \protect\cite{hires}. AMANDA UHE limit \protect\cite{amandauhe} and Tauwer foreseen $\nu_{\tau}$ sensitivity \protect\cite{tauwer} are for $E^{-2}$ fluxes. Models are from \protect\cite{allard} (pure proton and mixed compositions) and \protect\cite{engel01}. }
\label{fig8}
\end{figure}

Cosmogenic neutrinos are often considered as a guaranteed Ultra High Energy (UHE) flux but predicted fluxes vary over orders of magnitudes. Their detection may be challenging for existing detectors and new techniques may be needed to firmly establish their existence and to extract useful astrophysical information. 
Protons with energies $\gtrsim 10^{19}$ eV interact with the cosmic microwave and infrared backgrounds producing pions that decay into neutrinos. On the other hand, if the UHE cosmic rays (UHECRs) consist of heavy or intermediate mass nuclei rather than protons, they would generate neutrinos through photodisintegration followed by pion production through nucleon-photon scattering and resulting fluxes would be lower than in the pure proton assumption.  Different models for cosmogenic neutrinos have been described at this conference in Ref.~\cite{berezinsky,allard}. The various models span a large range due to the degree of freedom in defining the transition region between the galactic fading component and the onset of the extragalactic one and
in assumptions on UHECR composition and spectra since these measurements are affected by large systematics. The main parameters on which models depend are the source injection spectrum and its maximum energy, the source evolution and the source composition. Models are normalized to measurements by HiReS, Pierre Auger Observatory (PAO) and Akeno/AGASA. Fluxes normalized to PAO and HiReS at $10^{19}$ eV \cite{allard,luis} use a normalization of about a factor of 2 lower than those normalized to AGASA \cite{engel01}. The transition from galactic to extragalactic is assumed at energies around $10^{18}$ eV in the case of `dip models' \cite{berezinsky}, while 
`ankle models'  are normalized at $10^{19}$ eV and injection spectra can be harder resulting in higher neutrino fluxes, for similar assumptions on composition and source evolution \cite{allard}.
In Fig.~\ref{fig8} we show some of these models and compare them to current experimental results.

Various R\&D activities to measure noise levels and backgrounds for new techniques are ongoing at the South Pole: radio antennas, working in the region 0.2-1.2 GHz, similar in concept to Rice \cite{rice} and ANITA \cite{anita} that measure the Cherenkov radiation produced by the electromagnetic (EM) component of  neutrino induced showers due to the Askaryan effect \cite{rice}; surface radio antennas (20-250 MHz)  suitable for measuring the geo-synchrotron radio emission from cosmic ray showers, a technique pioneered by LOPES \cite{lopes}; acoustic detection with hydrophones of pulses (tens of kHz) due to fast thermal energy deposition of particle cascades. Ice has an attenuation length for the Cherenkov light of $\sim 100$ m, and the advantage of radio or acoustic techniques is a larger attenuation length (for radio $\sim 1$ km, for acoustic not yet better constrained than $>200$ m). 
Three clusters of radio transmitters and receivers where installed in IceCube holes at 1.4 km and 300 m and 3 others are going to be installed this season.
It was demonstrated that triggers are possible though the technique is not background free and more work on reconstruction techniques will help to discriminate man made backgrounds.


\begin{thebibliography}{19}
\bibitem{daq} R.~Abbasi {\it et al.}, arXiv:0810.4930, subm. to NIM.
\bibitem{first} A.~Achterberg {\it et al}, Astrop. Phys. 26 (2006) 155.
\bibitem{bazo} J.~Bazo for the IceCube Collaboration, these proceedings.
\bibitem{bartol} G.~Barr {\it et al}, Phys. Rev. D70 (2004) 023006. 
\bibitem{taup07} T.~Montaruli for the IceCube Collaboration, J. Phys. Conf. Ser. {\bf 120} (2008) 062009. 
\bibitem{amanda7} R.~Abbasi {\it et al.}, arXiv:0809.1646, subm.to Phys. Rev. D, (2008).
\bibitem{sk} K. Abe {\it et al.}, Astrophys. J. 652 (2006) 198.
\bibitem{macro} M.~Ambrosio {\it et al.}, Astrophys. J.  546 (2001) 1038.
\bibitem{berezinsky} V. Berezinsky, these proceedings.
\bibitem{allard} D. Allard, these proceedings and D. Allard {\it et al.}, JCAP 0609 (2006) 005.
\bibitem{luis} L.A. Anchordoqui {\it et al.}, Phys. Rev. D76 (2007) 123008.
\bibitem{engel01} R. Engel, D. Seckel and T. Stanev, Phys. Rev. D64 (2001) 093010.
\bibitem{anita} S.W. Barwick {\it et al.},  Phys. Rev. Lett. 96 (2006) 171101.
\bibitem{gorham} P~ Gorham, to appear in proc. of Neutrino 2008, Christchurch, New Zealand.
\bibitem{rice} L.~Kravchenko {\it et al.}, Phys. Rev. D73 (2006) 082002.
\bibitem{eheicrc07} A. Ishihara for the IceCube Coll., proc. of ICRC2007, astro-ph/0711.0353.
\bibitem{PAO} J. Abraham {\sl et al.}, Phys. Rev. Lett. 100 (2008) 211101.
\bibitem{hires} R.~Abbasi {\sl et al.}, arXiv:0803.0554.
\bibitem{amandauhe} M.~Ackermann {\it et al}, Astrop. J 675 (2008) 1014.
\bibitem{tauwer} M.~Iori {\sl et al.}, astro-ph/0602108.
\bibitem{lopes} H.~Falcke {\sl et al.}, Nature 435 (2005) 313.
\end{thebibliography}
\end{document}